\newif\ifjournal

\ifjournal
\documentclass[times,final]{elsarticle}
\usepackage{jcomp}
\usepackage{framed,multirow}
\else
\documentclass{scrartcl}
\fi

\usepackage[utf8x]{inputenc}
\usepackage{microtype} 

\usepackage[english]{babel} 

\usepackage{amsmath}
\usepackage{amssymb}

\usepackage{hyperref} 
\usepackage{graphicx}
\usepackage{booktabs}
\usepackage{pgf}
\usepackage{todonotes}

\usepackage{xcolor}
\definecolor{darkgreen}{RGB}{0,170,0}

\newcommand{\f}{f_q}
\newcommand{\intVel}[1]{\int #1 \; d \vel }
\newcommand{\vel}[0]{\pmb{\xi}}
\newcommand{\fvel}[0]{\pmb{u}}

\newcommand{\bx}[0]{\pmb{x}}
\newcommand{\bxi}[0]{\pmb{\xi}}
\newcommand{\bu}[0]{\pmb{u}}

\newcommand{\feq}{f_q^{(eq)}}

\newcommand{\x}{\pmb{x}}

\renewcommand{\c}{\pmb{c}}

\newcommand{\ua}{u_\alpha}
\newcommand{\ub}{u_\beta}
\newcommand{\ug}{u_\gamma}

\newcommand{\ca}{c_{q\alpha}}
\newcommand{\cb}{c_{q\beta}}
\newcommand{\cg}{c_{q\gamma}}

\newcommand{\contMom}[2]{\mathcal{M}_C \left[ #1, #2 \right] }
\newcommand{\eqMom}{\Pi^{(eq)}}

\newcommand{\bTens}[2]{ \pmb{#1}^{(#2)} }
\newcommand{\bTensS}[3]{ #1^{(#2)}_{\pmb{#3}} }

\ifjournal
\journal{Journal of Computational Physics}
\fi

\begin{document}

\ifjournal
\verso{M. Bauer, U.Rüde}

\begin{frontmatter}

\title{An improved lattice Boltzmann D3Q19 method based on an alternative equilibrium discretization}

\author[1]{Martin Bauer\corref{cor1}}
\cortext[cor1]{Corresponding author: 
  Tel.: +49-9131-85-27927;  
  E-mail: martin.bauer@fau.de}
\author[1,2]{Ulrich Rüde}

\address[1]{Chair for System Simulation, Friedrich–Alexander–Universität Erlangen–Nürnberg, Cauerstraße 11, 91058 Erlangen,Germany}
\address[2]{CERFACS, 42 Avenue Gaspard Coriolis, 31057 Toulouse Cedex 1, France}

\received{}
\finalform{}
\accepted{}
\availableonline{}
\communicated{}
\else
\title{An improved lattice Boltzmann D3Q19 method based on an alternative equilibrium discretization}
\author{Martin Bauer\footnote{martin.bauer@fau.de}, Ulrich Rüde}
\maketitle
\fi

\begin{abstract}
Lattice Boltzmann simulations of three-dimensional, isothermal hydrodynamics often use either the D3Q19 or the D3Q27 velocity sets. While both models correctly approximate Navier-Stokes in the continuum limit, the D3Q19 model is computationally less expensive but has some known deficiencies regarding Galilean invariance, especially for high Reynolds number flows. 
In this work we construct lattice Boltzmann equilibria for hydrodynamics directly from the continuous Maxwellian equilibrium.
While this approach reproduces the well known LBM equilibrium for D2Q9 and D3Q27 lattice models,
it yields a different equilibrium formulation for the D3Q19 stencil. This newly proposed formulation is shown to be more accurate than the widely used, second order equilibrium, while having the same computation costs.
We present a steady state Chapman-Enskog analysis of the standard and the improved D3Q19 model and conduct numerical experiments  
that demonstrate the superior accuracy of our newly developed D3Q19 equilibrium.
\end{abstract}

\ifjournal
\end{frontmatter}
\fi



\section{Introduction}

The lattice Boltzmann method is a promising alternative to established finite element or finite volume flow solvers due
to its suitability for modern, parallel computing hardware. 
While these established methods are based on a direct discretization of the Navier-Stokes equations (NSE), describing the fluid macroscopically, the lattice Boltzmann Method (LBM) is based on a mesoscale description of the fluid. 
The LBM is rooted in kinetic theory and operates on distribution functions in phase-space and thus requires, besides a space and time discretization, a discretization of the velocity space as well. 
Different discrete velocity sets (lattices) have been devised that all fulfill the necessary symmetry and isotropy conditions up to second order, to correctly capture the macroscopic behavior of the fluid. While for the simulation of isothermal hydrodynamics in three dimensions a stencil with $15$ velocities is sufficient, stencils with $19$, $27$ or more directions show better results in terms of isotropy and rotational invariance.
Despite the lower accuracy of the reduced D3Q15 and D3Q19 stencils, they are still
used in practice since they are computationally less costly. The performance of the widely used single- and two relaxation time models is limited by memory bandwidth, not by in-core execution times, on almost all current hardware architectures if the compute kernels are well optimized \cite{Wellein2006,Wittmann2013a}. The more entries a stencil has, the higher is the pressure on the memory interface, and the lower the overall runtime. The roofline performance model predicts in this case that the execution time is directly proportional to the number of neighboring values accessed, rendering the D3Q19 model by a factor of $1.4$ faster than the D3Q27 scheme, assuming equal resolution and number of time steps.
However, a lack of rotational invariance has been shown for the D3Q19 model in certain simulation setups.
Non-axisymmetrical solutions have been reported by \cite{harrison2007use} in a constricted, axisymmetric tube geometry 
which contradict the expected symmetry of the problem. Especially for turbulent flows, discrepancies between the
D3Q19 and the D3Q27 model are reported \cite{White2011,mayer2006direct,kang2013effect}.

In this paper we propose an improved D3Q19 model that uses a modified equilibrium formulation to mitigate some of the reported accuracy problems
of the standard D3Q19 model. The well-established second or third order LB equilibria can be derived either in a top-down or bottom-up way. 
In the top-down approach the equilibrium is constructed making a generic ansatz and conducting a Chapman-Enskog analysis that connects the mesoscopic 
LB description to a macroscopic partial differential equation (PDE). 
The coefficients of the generic ansatz are then chosen such that in the macroscopic limit the Navier-Stokes equations are approximated. Alternatively, in a bottom-up approach the discrete equilibrium can be constructed by discretizing the Boltzmann equation using Hermite polynomials \cite{He1997, Shan1998, Shan2006}. Both approaches lead to the same discrete equilibrium formulation. We propose a modified discrete equilibrium for the D3Q19 lattice to remedy anisotropic truncation errors. The idea to modify the equilibrium is not new. Other equilibrium modifications have been proposed as well to eliminate particular defects of the standard equilibrium \cite{ginzburg2008consistent}.
In this paper we first review the lattice Boltzmann equation and its bottom-up derivation from the continuous Boltzmann equation via projection on a Hermite subspace. Building on that, we then present our technique to construct an improved equilibrium for the D3Q19 model by matching moments of the continuous Maxwellian equilibrium. Both equilibrium formulations are then analyzed and compared via a steady state Chapman-Enskog analysis. To support the theoretical conclusions, numerical experiments are conducted that show how known shortcomings of the D3Q19 stencil - compared to D3Q27 - can be alleviated by our new equilibrium formulation.

\section{From the Boltzmann equation to Lattice Boltzmann via Hermite Projection}

Historically, the lattice Boltzmann method was developed as an improvement to lattice gas automata \cite{frisch1987lattice}. Lattice gas automata simulate the macroscopic behavior of fluids by constructing a discrete microscopic system that shows the same macroscopic behavior after sufficient coarse graining. Later it was shown in \cite{He1997} and \cite{Shan2006}, that the LB method can also be interpreted as a special discretization of the Boltzmann equation. In this section we review and summarize the process of deriving the single relaxation time (SRT) lattice Boltzmann method from the Boltzmann equation because our construction of an improved D3Q19 model presented in section \ref{sec:improvedEqConstruction} uses a similar approach.

\subsection{Boltzmann BGK}
The Boltzmann equation describes the time evolution of the probability density function $f(t, \bx, \bxi)$ of a single particle in phase space subject to an external force $F_i$
\begin{equation}
  \label{eq:boltzmann}
  \partial_t f + \xi_i \, \partial_i f + F_i \, \partial_{\xi_i} f = C(f).
\end{equation}
Partial derivatives with respect to spatial coordinates are abbreviated as $\partial_i$ and Einstein summation convention is used.
If derivatives w.r.t. Cartesian coordinates are written out explicitly, they are denoted as $\partial_x$, $\partial_y$, $\partial_z$ or $\partial_0$, $\partial_1$, $\partial_2$ using zero-based indexing. Time derivatives are written as $\frac{\partial}{\partial t} = \partial_t $. 
The probability distribution function $f(t, \bx, \bxi)$ represents the density of particles at time $t$ at position $\bx$ with velocity $\bxi$. The right hand side $C(f)$ denotes a generic collision term, constructed such that it conserves density, momentum and energy. These constraints can be expressed as 

\vspace{0.2cm}
\begin{minipage}{0.25\linewidth}
\begin{equation}
\label{eq:collMassCons}
\intVel{ C(f) } = 0
\tag{C1}
\end{equation}
\end{minipage}
\hspace{0.5cm}
\begin{minipage}{0.25\linewidth}
\begin{equation}
 \label{eq:collMomentumCons}
 \intVel{ \xi_i \, C(f) } = 0 
 \tag{C2}
\end{equation}
\end{minipage}
\hspace{0.5cm}
\begin{minipage}{0.25\linewidth}
\begin{equation}
 \label{eq:collMomentumCons}
 \intVel{ \xi_i \xi_i \, C(f) } = 0 
 \tag{C3}
\end{equation}
\end{minipage}
\vspace{0.35cm}

\noindent
with all integrals ranging over three dimensional velocity space. While energy conservation is used to derive the Maxwellian equilibrium, the LB methods considered in the following sections are not energy conserving. We introduce a shorthand notation for continuous velocity moments,

\begin{equation}
    \contMom{ P(\vel) } {f(\bxi)} := \intVel{P(\bxi) f(\bxi) }
\end{equation}
with $P(\bxi)$ being an arbitrary polynomial in $\bxi$. The conservation constraints on the collision operator (C1)-(C3) 
in this notation are expressed as $\contMom{P(\bxi)}{C(f)} = 0$  for  $P(\bxi) \in \{1, \xi_i, \xi_i \xi_i \} $.
Boltzmann's original collision term was derived assuming that the particle distributions before the collision are independent and uncorrelated (molecular chaos assumption). Boltzmann showed that a certain quantity~$H$, which is related to the entropy can only ever decrease ($H$-theorem). For an ideal gas, $H$ is the negative entropy density and thus the H-theorem is the kinetic theory basis for the second law of thermodynamics. The Maxwellian equilibrium distribution $\feq$ can be derived as a minimizer of $H$ under the constraints (C1)-(C3) and reads

\begin{equation}
\label{eq:maxwellian}
  f^{(eq)}(\rho, \bu) = \frac{\rho}{(2\pi \theta)^{\frac{D}{2}} }   \exp\left( -\frac{||\vel-\fvel||^2}{2\theta} \right).
\end{equation}
The BGK collision operator 
\begin{equation}
 C_{BGK}(f) := - \omega \left( f - f^{(eq)}(\rho,\bu) \right) 
\end{equation}
approximates Boltzmann's original collision term by linearly relaxing $f$ to a local equilibrium $f^{(eq)}$ with a constant rate $\omega$. 
The local equilibrium $f^{(eq)}$ is parametrized by the density $\rho=\contMom{1}{f}$ and the macroscopic velocity $\bu=\contMom{\bxi}{f}$. In the following we illustrate all concepts using the BGK collision term and show later how the results generalize to 
more advanced models, like multi-relaxation-time (MRT) or entropic collision operators.


\subsection{Hermite Expansion and Quadrature}
\label{sec:hermiteDerivation}

The LB equation can be derived as a projection of the Boltzmann equation on a discrete subspace, spanned by a set of orthogonal Hermite
polynomials. This particular subspace is chosen, because conservation properties of the collision operator are still exactly fulfilled after the projection. We show in this section, that knowing the first expansion coefficients of a Hermite-expanded function is equivalent to knowing the value of the distribution function at a set of discrete velocities. The connection between both representations are the Gauss-Hermite quadrature formulas. 

Following \cite{Shan2006}, we begin the LB derivation by introducing the d-dimensional Hermite polynomials
\begin{equation}
  \bTens{H}{n}(\bxi) = (-1)^n \frac{1}{W(\bxi)} \bTens{\nabla}{n} \; W(\bxi) 
\end{equation}
using the weight function 
\begin{equation}
  W(\bxi) = \frac{1}{(2\pi)^{D/2} } \exp\left(-\frac{\bxi^2}{2}\right) 
\end{equation}
and the abbreviation $ \nabla^{(n)}_{ij\cdots k} = \partial_{\xi_i} \partial_{\xi_j} \cdots \partial_{\xi_k} $. The tensor $\bTens{\nabla}{n}$ has $n$ indices, each representing a derivative with respect to a velocity coordinate.
Hermite polynomials fulfill the following orthogonality conditions and thus form an orthogonal basis with respect to the $W$-weighted scalar product $\left< f, g \right>_W = \int W f g \, d\vel$ 

\begin{equation}
\label{eq:hermite_orthogonality}
\int W(\bxi) \;\; \bTensS{H}{m}{i}(\bxi) \;\; \bTensS{H}{n}{j}(\bxi) \;\; d\bxi =
 \begin{cases}
    0  &  n \neq m \\
    \bTensS{\delta}{n}{ij} & n = m
 \end{cases}
\end{equation}
$\bTens{\delta}{n}_{\pmb{i}\pmb{j}}$ is the generalized Kronecker symbol which is $1$ iff 
the multi index $\pmb{i} = (i_1, i_2, ... i_n)$ is a permutation of $\pmb{j} = (j_1, j_2, ... j_n)$.
%
The distribution function $f$ can be expanded in the Hermite basis as 
\begin{equation}
\label{eq:hermite_expansion}
f(\vel) = W(\vel) \sum_{n=0}^\infty \frac{1}{n!} \bTensS{a}{n}{i}  \bTensS{H}{n}{i}(\vel)
\end{equation}
where the expansion coefficients $\bTensS{a}{n}{i}$ are linear combinations of moments of $f$.
%
%
\begin{equation}
 \bTensS{a}{n}{i} = \intVel{ f(\vel) \; \bTensS{H}{n}{i}(\vel)} = \contMom{\bTensS{H}{n}{i}}{f} 
\end{equation}
We project the Boltzmann equation \eqref{eq:boltzmann} on the Hermite subspace of order $N$ which discretizes the distribution function in the velocity coordinates.
\[   \tilde{f}(\vel) = W(\vel) \sum_{n=0}^N \frac{1}{n!} \bTensS{a}{n}{i}  \bTensS{H}{n}{i}(\vel)   \]
The Hermite basis is used for projection, because the approximation $\tilde{f}(\vel)$ of order $N$ matches the first $N$ moments of the approximated function $f$ exactly i.e. $\contMom{P(\bxi)}{\tilde{f}} = \contMom{P(\bxi)}{f}$ for polynomials $P(\xi)$ of degree not greater than $N$.
%
%
LB schemes do not use Hermite expansion coefficients to represent the distribution function because the propagation step is hard to formulate in this representation. Instead the evaluations of the projected distribution function $\tilde{f}(\vel_a)$ at a set of discrete velocities $\vel_a$ are stored. Both representations are connected via Gauss-Hermite quadrature formulas.
The 1D Gauss-Hermite quadrature rule approximates the integral of a univariate function $g(\xi)$
by a weighted sum of the function evaluated at a set of abscissae $\xi_a$
\[  \int W(\xi) g(\xi) \; dx \approx \sum\limits_{a=1}^{N} w_a g(\xi_a) .
\]
Above quadrature rule has algebraic accuracy of $(2N-1)$ meaning that if $g(\xi)$ is a polynomial of degree not greater than
$(2N-1)$, the approximation is exact. 
Our goal is to derive LB velocity sets that access the first neighborhood only. Since the entries in the LB velocity set are abscissae of the quadrature rule, we choose $N=3$, evaluating $g$ at the center and the direct left and right neighbors.
The choice of $N=3$ results in an algebraic accuracy of $5$.
The quadrature abscissae are the roots of $H^{(3)}(\xi)$, $ \xi_a =( -\sqrt{3}, 0, \sqrt{3} )$ 
and the corresponding weights are $w_i = \left(\frac{1}{6}, \frac{2}{3}, \frac{1}{6} \right)$.
For multivariate functions there is no general theory on Gauss quadrature. 
However, in case of Gauss-Hermite quadrature the multidimensional weight function $W(\bxi)$ can be factorized and 
the multidimensional integral can be written as a product of 1D integrals. 
This directly leads to the D2Q9 and D3Q27 velocity sets. Using symmetry arguments the smaller 
stencils D3Q15 and D3Q19 can be constructed \cite{Shan2006}, which still have an algebraic accuracy of $5$. 
To summarize, the third order Gauss-Hermite quadrature rule exactly integrates polynomials up to fifth order 
and leads to the discrete velocity sets D1Q3, D2Q9, D3Q15, D3Q19 and D3Q27 with abscissae magnitude $\sqrt{3}$. 
%

%
Using above quadrature rule we can express the coefficients $\bTensS{a}{n}{i}$ of the discrete distribution 
function $\tilde{f}(\vel)$ as a function of evaluations of $\tilde{f}(\vel_q)$ at discrete stencil velocities $\bxi_q \in S$. 
If $\tilde{f}(\bxi_q)$ is known for all abscissae $\vel_q$ of a quadrature rule with algebraic accuracy of at least $2N$ the expansion coefficients $\bTensS{a}{n}{i}$ for $n \le N$  can be obtained using

 \[ \bTensS{a}{n}{i} = \intVel{ W(\bxi) 
                                 \underbrace{ \frac{ \tilde{f}(\vel)}{W(\vel)}  }_{\substack{\text{\tiny Polynomial} \\ \text{\tiny of order $\le N$}}}
                                 \; 
                                 \underbrace{\bTensS{H}{n}{i}(\vel)}_{\substack{\text{\tiny Polynomial} \\ \text{\tiny of order $\le N$}}}
                               }
                     = \sum_{q} \frac{w_q}{W(\bxi_q) } \tilde{f}(\bxi_q) \bTensS{H}{n}{i}
                     = \sum_{q}  f_q \, \bTensS{H}{n}{i}
 \]
where the discrete distribution function $f_q$ is defined as $f_q := \frac{w_q}{W(\xi_q) } \tilde{f}(\xi_q)$.
In case of first neighborhood stencils where fifth order quadrature accuracy is obtained, the coefficients up to $\lfloor \frac{5}{2} \rfloor = 2$ can be represented exactly. 
Projecting the Boltzmann BGK equation \eqref{eq:boltzmann} on Hermite space and pointwise evaluation at the abscissae $\xi_q$ of the quadrature rule gives
\begin{equation}
 \label{eq:velocityDiscreteBoltzmann}
  \partial_t \tilde{f}(\xi_q) + \xi_i \partial_i \tilde{f}(\xi_q) = -\omega \left( \tilde{f}(\xi_q) - \tilde{f}^{(eq)}(\xi_q) \right) .
\end{equation}
Multiplying the projected equation with $\frac{w_q}{W(\xi_q) }$ leads to the LB equation \eqref{eq:lb_srt_with_forces}. The force terms have been left out here, but can be derived similarly~ \cite{Shan2006}.
If the Hermite expansion is truncated for $N\ge 2$, the moments up to second order of the approximation $\tilde{f}$ exactly coincide with the moments of $f$. Thus, the conservation properties of the collision operator are still satisfied by the projected equation if the equilibrium is also projected on the Hermite subspace.
%
%
%
%
Choosing the temperature parameter $\theta = 1$ in the Maxwellian equilibrium, projecting it on second (third) order Hermite subspace,
and rescaling velocities by $\sqrt{3}$ to obtain integer quadrature abscissae yields

\begin{equation} \label{eq:discreteEquilibriumStandard}
\tilde{f}^{(eq)}_q =  w_i \rho + w_i \rho_0  
                     \left[  3 
                     \ca u_{\alpha}
                              + \underbrace{  \frac{9}{2} \ua \ub (\ca \cb - \frac{1}{3} \delta_{\alpha\beta})  }_{\mbox{2nd order}}
                              + \underbrace{  \frac{9}{2} \left( \ua \ub \ug ( \ca \cb \cg - \frac{1}{3} \ca \delta_{\beta\gamma} )  \right) }_{\mbox{3rd order}}
                     \right]
\end{equation}
with $\rho_0=\rho$. This formulation approximates the weakly compressible NSE equations, and is often called the compressible LBM.
An equilibrium that better approximates incompressibility is obtained by choosing $\rho_0=1$ \cite{zou1995improved, he1997theory}. 
Note that the derived collision operator conserves mass and momentum but is not energy conserving.

In the supplementary material we provide an implementation of this Hermite projection approach in a compute algebra system together with further details on the derivation of~\eqref{eq:discreteEquilibriumStandard}.

\subsection{Lattice Boltzmann Model}

The following analysis and numeric test cases use the equilibrium version where $\rho_0=1$.
The velocity-discrete Boltzmann equation \eqref{eq:velocityDiscreteBoltzmann} is discretized in space and time on a regular lattice with cell size $\Delta x$ and a discrete time step $\Delta t$. 
%
\begin{equation}
 \f(\x + \vel_q \, \Delta t, t) = \f(\x, t) - \omega (\f - \feq)  + \Delta x \left( 1 - \frac{\omega}{2} \right)  F_q(\x, t)
 \label{eq:lb_srt_with_forces}
\end{equation}
The velocities $\vel_q$ are connected to the integer lattice velocities $c_q$ via the scaling $\vel_q = c \, \c_q$, where $c = (\Delta x / \Delta t)$. 
$F_q$ is a forcing term defined as \cite{Buick2000}
\[ F_q(\x, t) = \frac{w_q }{c_s^2} \, \c_q \cdot \pmb{a} \]
with lattice dependent weights $w_q$. We deliberately use a forcing scheme with vanishing second order moments instead of \cite{Guo2002} since
we want to approximate the incompressible Navier Stokes equation in steady state~\cite{Kruger2016, Silva2011}. 
Similar to the continuous case, macroscopic quantities are obtained as moments of the distribution function. 
Note the force dependent shift of the macroscopic velocity that also enters the equilibrium distribution.
\begin{equation}
\rho = \sum_q \f =  \sum_q \f^{(eq)}
\label{eq:mass_conservation}
\end{equation}
\begin{equation}
 \pmb{u} = \sum_q \vel_q \f + \frac{\Delta x}{2} \sum_q \vel_q F_q = \sum_q \vel_q \f^{(eq)} 
 \label{eq:macroscopic_vel_shift}
\end{equation}
The SRT/BGK collision model was chosen instead of more elaborate collision operators because the collision operator does not change the rotational invariance behavior and the theoretical analysis is considerably simpler for the SRT case.
\vspace{0.6cm}
\\
\textbf{D3Q19}:
\[
\bxi_q = c
\begin{cases}
   (0,0,0)                                                 & q  = 0,\\
   (\pm 1, 0, 0), (0, \pm 1, 0), (0, 0, \pm1)              & q = 1, 2, ... , 6 \\
   (\pm 1, \pm 1, 0), (\pm 1, 0 \pm 1,), (0, \pm 1, \pm 1) & q = 7, 8, ... , 18
\end{cases}
\]
\textbf{D3Q27}
\[
\bxi_q = c
\begin{cases}
   (0,0,0)                                                 & q  = 0,\\
   (\pm 1, 0, 0), (0, \pm 1, 0), (0, 0, \pm1)              & q = 1, 2, ... , 6 \\
   (\pm 1, \pm 1, 0), (\pm 1, 0 \pm 1,), (0, \pm 1, \pm 1) & q = 7, 8, ... , 18 \\
   (\pm 1, \pm 1, \pm 1)                                   & q = 19, 8, ... , 26
\end{cases}
\]
with weights for D3Q19 stencil being
\[ w_q = \begin{cases}
	1/3  & q =  0 \\ 
	1/18 & q = 1, 2, ... , 6 \\
    1/36 & q = 7, 8, ... , 18
\end{cases}	
\]
and for D3Q27
\[ w_q = \begin{cases}
   8/27   & q  = 0,\\
   2/27   & q = 1, 2, ... , 6 \\
   1/54   & q = 7, 8, ... , 18 \\
   1/216  & q = 19, 8, ... , 26
\end{cases}	
\]
%

\section{Construction of improved D3Q19 equilibria}
\label{sec:improvedEqConstruction}

\subsection{Moment matching approach}

In this section we use an alternative technique to construct a discrete equilibrium formulation using directly the continuous Maxwellian distribution.
The Hermite projection approach presented in section \ref{sec:hermiteDerivation} ensures that the moments up to second order of the discrete equilibrium match their continuous counterparts if a Gauss-Hermite quadrature rule of at least algebraic order $4$ is used. This is the case for the commonly used single speed velocity sets D2Q9, D3Q15, D3Q19 and D3Q27.
We obtained the lattice Boltzmann equation as a discretization of the continuous Boltzmann equation. 
It is not self evident that the LB equation, obtained in this way, approximates the macroscopic Navier-Stokes equations.
This can be shown through a Chapman Enskog analysis which is described in detail in \cite{Kruger2016}.
In the limit of small grid spacings and low Mach number the NSE are obtained if the equilibrium moments satisfy the conditions
\begin{subequations} \label{eq:equilibrium_moment_conditions}
\begin{alignat}{3}
\Pi^{(eq)}       = & \rho  \\
\Pi^{(eq)}_i     = & \rho_0 u_i \\
\Pi^{(eq)}_{ij}  = & \rho_0 u_i u_j + \delta_{ij} (c_s^2 \rho)  \\ 
\Pi^{(eq)}_{ijk} = & \rho_0 c_s^2 ( u_i \delta_{jk} + u_j \delta_{ik} + u_k \delta_{ij}  ) + \mathcal{O}(u^3).
\end{alignat}
\end{subequations}
We use the shorthand notation $\Pi_{i_1 i_2 \cdots i_n}^{(eq)} = \sum_q \xi_{q\, i_1} \xi_{q\, i_2} \cdots \xi_{q\, i_n} f^{(eq)} $ to denote equilibrium moments.
Again, for $\rho_0=\rho$ the weakly compressible NSE are obtained, for $\rho_0=1$ the incompressible version is approximated.
Not surprisingly, the continuous Maxwellian equilibrium \eqref{eq:maxwellian} fulfills all these conditions in the compressible case. Thus, if the discrete equilibrium moments up to order 3 match the corresponding continuous moments of the Maxwellian, the LB scheme with this equilibrium approximates the compressible NSE.
The conditions \eqref{eq:equilibrium_moment_conditions} only require that the continuous moments are matched up to $\mathcal{O}(u^3)$.
The Hermite construction ensures that conditions (\ref{eq:equilibrium_moment_conditions}a-c) are fulfilled, however condition (\ref{eq:equilibrium_moment_conditions}d) is also fulfilled up to second order in $u$, as can be seen in Table \ref{tbl:moment_comparison}.

The central idea of the alternative bottom-up equilibrium construction approach is to build a discrete equilibrium that matches as many moments of the Maxwellian equilibrium \eqref{eq:maxwellian} as possible. With single speed stencils the number of different moments that can be represented in a discrete setting is limited by an effect called moment aliasing. Computing moments in lattice units where $c = 1$ and $\bxi_q = \c_q$, the discrete velocity components $\xi_{qi}$ can only take on values in $\{-1, 0, 1\}$. Thus, moments with velocity powers larger than 2 alias a lower order moment, e.g.
\begin{equation}
\Pi^{(eq)}_{iii} = \sum\limits_{\bxi_q \in S} \xi_{qi}^3 \, f^{(eq)}_q = \sum\limits_{\bxi_q \in S} \xi_{qi}^1 \, f^{(eq)}_q = \Pi^{(eq)}_i \;\; \;\; \mbox{if} \;\; \xi_{qi} \in \{-1, 0, 1\}  
\end{equation}
That means that for single speed velocity sets, $\Pi^{(eq)}_{iii}$ and $\Pi^{(eq)}_{i}$ cannot be controlled independently.
In general, for a $d$ dimensional, single speed stencil, only the moments $\xi_{q0}^{e_0} \cdot \xi_{q1}^{e_1} \cdots \xi_{qd}^{e_d}$ with exponents $e_i \in \{0, 1, 2\}$ are independent. For two dimensional models there are $9$, for 3D stencils $27$ moments that can be controlled independently. 

The D2Q9 and D3Q27 models are special in the sense that their number of discrete velocities is equal to the number of maximally possible
independent moments for single speed models.
To obtain a discrete equilibrium $f_q^{(eq)}$, we require all independent 9 or 27 moments to match their Maxwellian counterparts. The resulting linear system of equations can be uniquely solved for $f_q^{(eq)}$. For these full neighborhood stencils, exactly the same equilibrium is obtained as was derived via the Hermite projection \eqref{eq:discreteEquilibriumStandard} up to third order in $\bu$. If the Maxwellian moments are first truncated to second order in $\bu$, the commonly used second order equilibrium is obtained. 
Thus, our equilibrium construction technique yields the same equilibrium \eqref{eq:discreteEquilibriumStandard} as the Hermite projection for the D2Q9 and D3Q27 stencils.

In case of the D3Q19 model, however, there are only 19 degrees of freedom available to match 27 independent moments. 
Table \ref{tbl:moment_comparison} shows the Maxwellian moments and the equilibrium moments of the standard equilibrium evaluated on the D3Q19 stencil in the fourth and fifth column. Terms that are different are highlighted in bold. Only the first 10 moments are matching. If we neglect differences in $\mathcal{O}(u^3)$, $17$ moments are matching. These are exactly the conditions \eqref{eq:equilibrium_moment_conditions} required for the LB scheme to approximate the NSE. Since the standard equilibrium only matches $10$ (or $17$) moments, but has $19$ degrees of freedom, 
a new equilibrium can be constructed that matches $19$ Maxwellian moments. We label this new {\em improved} equilibrium D3Q19-I, in contrast to the {\em standard} D3Q19-S equilibrium. In the next subsection we describe the construction in detail using a formalism used in multi-relaxation-time (MRT) schemes and give an explicit formula for D3Q19-I.

\begin{table}
	\resizebox{\textwidth}{!}{%
	\begin{tabular}{llll|lll}
	 \# & Moment & Order & Maxwellian  & D3Q19-S & D3Q19-I & D3Q27 \\ 
	\hline
	1 & $\Pi^{(eq)}$                 & 0  & $\rho$ 
								& $\rho$ 
								& $\rho$ 
								& $\rho$ \\
	3 & $\Pi^{(eq)}_{i}$         & 1  & $\rho_0 u_i$  
 								& $\rho_0 u_i$  
								& $\rho_0 u_i$  
								& $\rho_0 u_i$  \\
	3 & $\Pi^{(eq)}_{ii} $      & 2  & $\rho_0 u_i^2 + c_s^2 \rho$ 
								& $\rho_0 u_i^2 + c_s^2 \rho$ 
								& $\rho_0 u_i^2 + c_s^2 \rho$ 
								& $\rho_0 u_i^2 + c_s^2 \rho$ \\
	3 &$\Pi^{(eq)}_{ij}$         & 2  & $\rho_0 u_i u_j$
        							& $\rho_0 u_i u_j$
								& $\rho_0 u_i u_j$
								& $\rho_0 u_i u_j$\\
	6 & $\Pi^{(eq)}_{iij}$ & 3  & $\frac{\rho_0}{3} u_j + \rho_0 u_i^2 u_j$  
	 							& $\frac{\rho_0}{3} u_j + \rho_0 u_i^2 u_j \mathbf{- \frac{\rho_0}{2} u_j u_k^2} $    
	 							& $\frac{\rho_0}{3} u_j + \rho_0 u_i^2 u_j$  
	 							& $\frac{\rho_0}{3} u_j + \rho_0 u_i^2 u_j$  	\\
	1 & $\Pi^{(eq)}_{ijk}$     & 3  & $\rho_0 u_i u_j u_k$  
	 							& $\mathbf{0}$    
	 							& $\mathbf{0}$ 
	 							& $\rho_0 u_i u_j u_k$  	\\		
	\hline	 									
	3 & $\Pi^{(eq)}_{iijj}$     & 4  & $\frac{\rho}{9} + \frac{\rho_0}{3}(u_i^2 + u_j^2)$
							    & $\frac{\rho}{9} + \frac{\rho_0}{3}(u_i^2 + u_j^2) \mathbf{- \frac{\rho_0}{6} u_k^2}$  
							    & $\frac{\rho}{9} + \frac{\rho_0}{3}(u_i^2 + u_j^2)$
							    & $\frac{\rho}{9} + \frac{\rho_0}{3}(u_i^2 + u_j^2)$\\
    3 &  $\Pi^{(eq)}_{iijk}$     & 4  & $\frac{\rho_0}{3} u_i u_j$ 
    								& $\mathbf{0}$
    								& $\mathbf{0}$
    								& $\frac{\rho_0}{3} u_i u_j$  \\
    	3 &  $\Pi^{(eq)}_{iijjk}$    & 5 & $\frac{\rho_0}{3}( \frac{u_k}{3} + u_i^2 u_k + u_j^2 u_k )$ 
    								& $\mathbf{0}$
    								& $\mathbf{0}$
    								& $\frac{\rho_0}{3}( \frac{u_k}{3} + u_i^2 u_k + u_j^2 u_k )$  \\ 
    1 & $\Pi^{(eq)}_{iijjkk}$    & 6  & $\frac{\rho_0}{9} (u_i^2 + u_j^2 + u_k^2) + \frac{\rho_0}{27}$ 
    								& $\mathbf{0}$
    								& $\mathbf{0}$
    								& $\frac{\rho_0}{9} (u_i^2 + u_j^2 + u_k^2) + \frac{\rho_0}{27}$ 
	\end{tabular}
	}
\caption{\footnotesize Overview of all non-aliased 3D equilibrium moments with $c_s^2=1/3$. The moments (second column) are written in shorthand notation to summarize multiple moments into a single row, with $i,j,k \in \{0,1,2\}$ and $i \neq j \neq k$. This notation is different to usual notation: note that index variables may not have the same value and repeating indices are not summed over. For example the abbreviation $\Pi^{(eq)}_{iijjk}$ summarizes the moments $\xi_0^2 \xi_1^2 \xi_2, \xi_1^2 \xi_2^2 \xi_0, \xi_0^2 \xi_2^2 \xi_1$. 
The number of summarized moments in a row is given by the first column.
Moments up to order $3$ have to match the Maxwellian moments up to $\mathcal{O}(u^2)$ to approximate the NSE.
The D3Q27 equilibrium matches all $27$ non-aliased moments.
The established D3Q19-S model matches $17$ moments up to $\mathcal{O}(u^2)$ and 10 moments up to $\mathcal{O}(u^3)$. The D3Q19-I equilibrium matches $20$ moments of the continuous Maxwellian up to $\mathcal{O}(u^2)$ and 19 moments up to $\mathcal{O}(u^3)$.
}
\label{tbl:moment_comparison}
\end{table}

\subsection{Derivation of D3Q19-I using MRT formalism}

The multi-relaxation-time (MRT) formalism \cite{Lallemand2000} provides a straightforward way to incorporate the moment constraints in the collision operator.
\begin{equation}
C_{MRT} \left[ \f(\x, t)  \right] = -M^{-1} S \left[Mf - m^{(eq)}  \right]  
\end{equation}
In the MRT collision process, the distribution functions are transformed to moment space first. The moment space is spanned by 
a set of $Q$ independent moments. The moment matrix $M$ maps a vector of distribution functions to a vector of moments. These
moments can then be relaxed with potentially different relaxation times through a diagonal matrix $S$ against equilibrium moments
$m^{(eq)}$. Usually the equilibrium moments are computed from the discrete equilibrium distribution \eqref{eq:discreteEquilibriumStandard} via $m^{(eq)} = M f^{(eq)}$. 
In our approach we directly use the moments of the Maxwellian \eqref{eq:maxwellian}.
For a $DdQq$ stencil, $q$ independent moments are required to fully define the discrete equilibrium.
The first neighborhood stencils that include all diagonals ($D2Q9$ and D3Q27) have $3^d$ discrete velocities.
For these models all $3^d$ linearly independent moments $\xi_{q0}^{e_0} \cdot \xi_{q1}^{e_1} \cdots \xi_{qd}^{e_d}$ with 
$e_i \in \{0, 1, 2\}$ are chosen. This construction yields an invertible moment matrix $M$. 
Applying our equilibrium construction approach for these stencils, gives exactly the equilibrium formulation as obtained by the Hermite projection method \eqref{eq:discreteEquilibriumStandard}. 
The D3Q19 stencil does not contain the 8 diagonal directions $(\pm 1, \pm 1, \pm 1)$ and has only $19$ discrete velocities. 
Accordingly only $19$ of the $27$ possible moments can be matched. To obtain the D3Q19-I equilibrium we drop the moments
\begin{equation}
\begin{split}
\xi_{q0}^{e_1} \xi_{q1}^{e_2} \xi_{q2}^{e_3}: (e_1, e_2, e_3) \in \{ & (1,2,2), (2,1,2), (2,2,1) \\
													 & (1,1,2), (1,2,1), (2,1,1) \\
													 & (1,1,1), (2,2,2) \}
\end{split}
\label{eq:dropped_d3q19_moments}
\end{equation}
The resulting moment space is the same as used in D3Q19 MRT methods \cite{DHumieres2002, Stiebler2011, Tolke2006, Khirevich2015}.
In contrast to these works we determine the equilibrium moments $m^{(eq)}$ from the continuous Maxwellian \eqref{eq:maxwellian} instead of the Hermite projected discrete equilibrium \eqref{eq:discreteEquilibriumStandard}.
Transformed back from moment space to physical space with $M^{-1}$, the D3Q19-I equilibrium truncated at velocity order 2 reads
\begin{equation}
f^{(eq)}_q = \rho w_q + w_q \, \rho_0 \cdot 
\begin{cases}
    - u_i u_i    &  \mathbf{c_q} = (0,0,0) \\
     3 u_i c_{qi} - 3 u_i u_i + 6 (u_i c_{qi} )^2 &  \mathbf{c_q} \in \{ (\pm1, 0,0), (0, \pm1, 0), (0,0,\pm1) \} \\
     3 u_i c_{qi} - \frac{3}{2} u_i^2 c_{qi}^2
                     + \frac{9}{2} (c_{qi} u_i)^2     & \mbox{else}
\end{cases}
\label{eq:new_d3q19_equilibrium}
\end{equation}
For further details on the D3Q19-I construction we refer to the supplementary material.

\subsection{Generalization to Advanced Collision Models}

Our improved D3Q19-I model is not restricted to the BGK/SRT collision operator. As previously shown, our equilibrium construction approach is more naturally expressed in moment space and thus straightforwardly generalizes to two relaxation time or multi relaxation time models. Instead of computing equilibrium moments $m^{(eq)}$ from a discrete equilibrium, they are obtained directly from the Maxwellian \eqref{eq:maxwellian}.
Entropically stabilized collision operators proposed by \cite{Bosch2015} improve MRT models by locally computing relaxation rates of higher order moments subject to a maximum entropy condition. Since these \texttt{KBC}-type collision operators change only relaxation rates, not equilibrium moments these methods can also be adapted to use Maxwellian moments. We show in section \ref{sec:numerical_experiments} that the D3Q19 KBC collision operator with Maxwellian moments gives more accurate results when adapted in this way.



\section{Chapman Enskog Analysis of D3Q19 Equilibria}
\label{sec:chapmanEnskog}

In this section we conduct a steady state Chapman Enskog analysis to determine and compare the truncation errors of the D3Q27, D3Q19-S and 
D3Q19-I models.
In the first step of the analysis, we introduce the scaling parameter $\epsilon$ and show that the scaled LB equation approximates the incompressible NSE in the limit $\epsilon \rightarrow 0$. We scale the equation diffusively with $\epsilon^2 = \Delta x^2 = \Delta t$,  so doubling the resolution leads to a time step reduced by a factor of 4. 
\begin{equation}
 \f(\x + \epsilon \, \c_q, t) = \f(\x, t) - \omega (\f - \feq) + \epsilon \left( 1 - \frac{\omega}{2} \right) F_q
 \label{eq:scaled_lb}
\end{equation}
For the steady state analysis of \eqref{eq:scaled_lb} we drop the time dependencies and expand the left hand side
in a Taylor series. 
\begin{equation}
\begin{split}
  \f(\x + \epsilon \, \c_q) = & \sum_{n=0}^\infty  \frac{1}{n!} \left( \epsilon \; c_{qi} \partial_i \right)^n \f(\x) \\
  	 = &
	 \left[ 
	  1  + 
	 \epsilon \left(   c_{q0} \partial_0  + c_{q1} \partial_1  \right) +
	 \frac{\epsilon^2}{2} \left( c_{q0}^2 \partial_0 \partial_0 + c_{q0} c_{q1} \partial_0 \partial_1 + c_{q1}^2 \partial_1 \partial_1   \right)+
	 \dots
	 \right] \f(\x)
\end{split}
\label{eq:taylor_expanded_lb}
\end{equation}
Spatial derivatives in $x$, $y$ and $z$ direction are denoted as $\partial_0$, $\partial_1$ and $\partial_2$ respectively.
Next, we assume that the solution $\f$ can be written as a power series \cite{Chapman1970} in the scaling parameter $\epsilon$
\begin{equation}
\f = \sum_{n=0}^\infty \epsilon^n \f^{(n)} = f^{(0)} + \epsilon f^{(1)} + \epsilon^2 f^{(2)} + \epsilon^3 f^{(3)} + \dots
\label{eq:chapman_enskog_ansatz}
\end{equation} 
Inserting equations~\eqref{eq:taylor_expanded_lb} and~\eqref{eq:chapman_enskog_ansatz} into~\eqref{eq:scaled_lb} and collecting terms with the same power in $\epsilon$ yields:
\begin{subequations} \label{eq:f_hierarchy}
\begin{alignat}{5}
f_q^{(0)} = & & f^{(eq)} \\
f_q^{(1)} =& -& \omega^{-1} &                            \;\; & c_{qi}&                     \;\;&\partial_i & f_q^{(eq)}                                         &+& \Lambda F_q   \\
f_q^{(2)} =&  &\omega^{-1} & \Lambda                     \;\; & c_{qi}& c_{qj}  \;\;& \partial_i & \partial_j f_q^{(eq)}                                         &-& \omega^{-1} \Lambda \partial_i c_{qi} F_q\\
f_q^{(3)} =& -& \omega^{-1}& \left( \Lambda^2 - \frac{1}{12}\right)\;\; & c_{qi}& c_{qj} c_{qk} \;\;&  \partial_i & \partial_j \partial_k f_q^{(eq)}                        &+& \Lambda^2  \partial_i \partial_j c_{qi} c_{qj} F_q   
\end{alignat}
\end{subequations}
For better readability the parameter $\Lambda=\omega^{-1} - \frac{1}{2}$ was introduced. 
Having determined the expansion terms $f^{(i)}$ separately, they can be recombined again to obtain the mass and momentum conservation equations. To get the mass conservation equation we make use of the property that the density is the zeroth moment of $f$ and the equilibrium distribution $f^{(eq)}=f^{(0)}$, and thus is conserved $ \rho = \sum_q f_q = \sum_q f_q^{(eq)} $.
Subtracting the equilibrium moment on both sides shows that the zeroth moment of the non-equilibrium part has to be zero
\begin{equation}
\sum_q (f_q - f_q^{(eq)}) = \sum_q \left( \epsilon f_q^{(1)} + \epsilon^2 f_q^{(2)} + \epsilon^3 f_q^{(3)} + \dots \right)  = 0.
\label{eq:recombination}
\end{equation}
Inserting \eqref{eq:f_hierarchy} into \eqref{eq:recombination} and canceling one factor of $\epsilon$ and $\omega^{-1}$ leads to an equation containing only equilibrium moments $\Pi^{(eq)}$ and $\Lambda$
\begin{equation}
- \partial_i \eqMom_i + 
\epsilon \Lambda  \left( \partial_i \partial_j \eqMom_{ij} - \partial_i a_i \right)
- \epsilon^2 \left( \Lambda^2 - \frac{1}{12} \right) \partial_i \partial_j \partial_k \eqMom_{ijk} 
= 0.
\label{eq:continuity_eq_moments}
\end{equation}
For the momentum conservation equation a similar approach is used. Additionally, here the shifting of the equilibrium velocity 
\eqref{eq:macroscopic_vel_shift} must be considered
\begin{equation}
   u_h = \sum_q c_{qh} \, f_q + \frac{\epsilon}{2} \sum_q c_{qh} F_q = \sum_q c_{qh} f_q^{(eq)} 
\end{equation}
\begin{equation}
 \sum_q c_{qh} \left( \epsilon f_q^{(1)} + \epsilon^2 f_q^{(2)} + \epsilon^3 f_q^{(3)} + \dots \right)  = - \frac{\epsilon}{2} \sum_q c_{qh} F_q . 
\end{equation}
Inserting the expanded distribution functions and expressing the terms as equilibrium moments yields
\begin{equation}
 \partial_i \eqMom_{ih} -
\epsilon \Lambda  \partial_i \partial_j \eqMom_{ijh} 
+ \epsilon^2 \left( \Lambda^2 - \frac{1}{12} \right) \partial_i \partial_j \partial_k \eqMom_{ijkh} 
=  a_h.
\label{eq:momentum_eq_moments}
\end{equation}
Additionally we used the fact that the force term $F_q$ has only first order moments that are non-zero. These non-zero first order moments are $\sum_q F_q c_{qh} = a_h$.
Equations \eqref{eq:continuity_eq_moments} and \eqref{eq:momentum_eq_moments} are independent of the used LB model.
Stencil and equilibrium specific equations are obtained by inserting equilibrium moments that have different values for different lattice models. Next, the moments of the incompressible equilibrium, truncated at second velocity order $\mathcal{O}(u^3)$, are inserted.
%
Since the D3Q27, D3Q19-S and D3Q19-I have the same equilibrium moments $\Pi_i, \Pi_{ij}$ and $\Pi_{ijk}$ up to second velocity order,
we get the same continuity equation for all three models by inserting their equilibrium moments into  \eqref{eq:continuity_eq_moments}.
\begin{equation}
\partial_i u_i = \epsilon \Lambda \partial_i \left(  \partial_j \left( \frac{\rho}{3} \delta_{ij} + u_i u_j   \right) - a_i  \right)
				 + \epsilon^2 \left( \Lambda^2 - \frac{1}{12}   \right) \partial_i \partial_j \partial_j u_i
\end{equation}
The left hand side is the macroscopic continuity equation for an incompressible fluid and the right hand side are error terms scaled by $\epsilon$. 
Similarly, the momentum transport equation is obtained by inserting equilibrium moments into \eqref{eq:momentum_eq_moments}. 
\begin{equation}
\partial_i \left( \frac{\rho}{3} \delta_{ih} + u_i u_h   \right) - a_h 
- \frac{1}{3} \epsilon \Lambda \left( \partial_i \partial_i u_h  + 2 \partial_i \partial_h u_i \right) 
= - \epsilon^2 \left( \Lambda^2 - \frac{1}{12} \right) \left( \partial_i \partial_i \partial_h \left( \frac{\rho}{3} \right) + E_h \right)
\end{equation}
Different error terms are obtained for the different lattice models, since forth order moments are entering the equation which have different values for D3Q27, D3Q19-S and D3Q19-I.
These error terms are

\begin{equation}
 E_h^{\text{D3Q27}} = 
		 \partial_i \partial_i \partial_j (u_h u_j) 
		 + \partial_i \partial_j \partial_h (u_i u_j) 
		  - \partial_h \partial_h \partial_h (u_h^2) 
\end{equation}
for the D3Q27 model, 
\begin{equation}
E_h^{\text{D3Q19-I}} = 
	 \partial_h \partial_i \partial_i ( u_h^2)
	+ 3 \partial_h \partial_h \partial_i ( u_h u_i)
	+ \delta_{ij} \partial_i \partial_i \partial_j (u_j u_h) 
	+ \delta_{ij} \partial_h \partial_i \partial_i (u_j u_j) 
    - 5 \partial_h \partial_h \partial_h (u_h^2)  
\end{equation}
for the newly proposed D3Q19-I model and
\begin{equation}
\begin{split}
E_h^{\text{D3Q19-S}} = 
	\partial_h \partial_i \partial_i (u_h^2)  
	+ 3 \partial_h \partial_h \partial_i (u_h u_i) 
	+ \delta_{ij} \partial_i \partial_i \partial_j (u_j u_h) 
	+ \delta_{ij} \partial_h \partial_i \partial_i (u_j u_j)  
    - 5 \partial_h \partial_h \partial_h (u_h^2)  \\
    - \frac{1}{2}(1 - \delta_{ij}) (1 - \delta_{ih})  (1 - \delta_{jh}) \partial_i \partial_i \partial_h (u_j^2)
\end{split}
\end{equation}
for the standard D3Q19-S equilibrium. In above equations, Einstein summation convention only applies to indices $i$ and $j$, not to the free index $h$.
In \cite{Silva2014} a similar analysis is conducted, however we obtain conflicting results for the second order errors $E_2$ compared to the work by Silva et al. Our analysis was done fully in a computer algebra system. For a detailed derivation, including the complete source code, we refer to the supplementary material.  

To better illustrate the difference in these error terms for different models, we consider a scenario that is invariant in $x$-direction, as is the case in the numerical experiments described in the next section. In the following equations, where no Einstein summation convention is used, we label the cartesian coordinates $x$, $y$ and $z$ using the indices $0$, $1$ and $2$. Writing out the second order error terms of the momentum transport equation $E_2$ for the different models after setting all $x$-derivatives to zero yields
\ifjournal \else \begin{footnotesize} \fi
\begin{equation}
E_2^{D3Q27}\bigg\rvert_{\partial_0 (\cdot) = 0} =
\left[\begin{matrix}{\partial_{1} {\partial_{1} {\partial_{1} (u_{0} u_{1}) }}} + {\partial_{1} {\partial_{2} {\partial_{2} (u_{0} u_{1}) }}} + {\partial_{1} {\partial_{1} {\partial_{2} (u_{0} u_{2}) }}} + {\partial_{2} {\partial_{2} {\partial_{2} (u_{0} u_{2}) }}}\\{\partial_{1} {\partial_{1} {\partial_{1} (u_{1}^{2}) }}} + {\partial_{1} {\partial_{2} {\partial_{2} (u_{1}^{2}) }}} + {\partial_{1} {\partial_{2} {\partial_{2} (u_{2}^{2}) }}} + 3 {\partial_{1} {\partial_{1} {\partial_{2} (u_{1} u_{2}) }}} + {\partial_{2} {\partial_{2} {\partial_{2} (u_{1} u_{2}) }}}\\{\partial_{1} {\partial_{1} {\partial_{2} (u_{1}^{2}) }}} + {\partial_{1} {\partial_{1} {\partial_{2} (u_{2}^{2}) }}} + {\partial_{2} {\partial_{2} {\partial_{2} (u_{2}^{2}) }}} + {\partial_{1} {\partial_{1} {\partial_{1} (u_{1} u_{2}) }}} + 3 {\partial_{1} {\partial_{2} {\partial_{2} (u_{1} u_{2}) }}}\end{matrix}\right]
\label{eq:errorTerms3D_D3Q27}
\end{equation}
\ifjournal \else \end{footnotesize} \fi
for the D3Q27 model,  
\ifjournal \else \begin{footnotesize} \fi
\begin{equation}
E_2^{\text{D3Q19-I}}\bigg\rvert_{\partial_0 (\cdot) = 0} =
\left[\begin{matrix}{\partial_{1} {\partial_{1} {\partial_{1} (u_{0} u_{1}) }}} + {\partial_{2} {\partial_{2} {\partial_{2} (u_{0} u_{2}) }}}\\{\partial_{1} {\partial_{1} {\partial_{1} (u_{1}^{2}) }}} + {\partial_{1} {\partial_{2} {\partial_{2} (u_{1}^{2}) }}} + {\partial_{1} {\partial_{2} {\partial_{2} (u_{2}^{2}) }}} + 3 {\partial_{1} {\partial_{1} {\partial_{2} (u_{1} u_{2}) }}} + {\partial_{2} {\partial_{2} {\partial_{2} (u_{1} u_{2}) }}}\\{\partial_{1} {\partial_{1} {\partial_{2} (u_{1}^{2}) }}} + {\partial_{1} {\partial_{1} {\partial_{2} (u_{2}^{2}) }}} + {\partial_{2} {\partial_{2} {\partial_{2} (u_{2}^{2}) }}} + {\partial_{1} {\partial_{1} {\partial_{1} (u_{1} u_{2}) }}} + 3 {\partial_{1} {\partial_{2} {\partial_{2} (u_{1} u_{2}) }}}\end{matrix}\right]
\label{eq:errorTerms3D_D3Q19I}
\end{equation}
\ifjournal \else \end{footnotesize} \fi
for the improved D3Q19-I model, and 
\ifjournal \else \begin{footnotesize} \fi
\begin{equation}
E_2^{\text{D3Q19-S}}\bigg\rvert_{\partial_0 (\cdot) = 0} =
\left[\begin{matrix}{\partial_{1} {\partial_{1} {\partial_{1} (u_{0} u_{1}) }}} + {\partial_{2} {\partial_{2} {\partial_{2} (u_{0} u_{2}) }}}\\
- \frac{1}{2} {\partial_{1} {\partial_{2} {\partial_{2} (\mathbf{u_{0}^{2}}) }}} + {\partial_{1} {\partial_{1} {\partial_{1} (u_{1}^{2}) }}} + {\partial_{1} {\partial_{2} {\partial_{2} (u_{1}^{2}) }}} + {\partial_{1} {\partial_{2} {\partial_{2} (u_{2}^{2}) }}} + 3 {\partial_{1} {\partial_{1} {\partial_{2} (u_{1} u_{2}) }}} + {\partial_{2} {\partial_{2} {\partial_{2} (u_{1} u_{2}) }}}\\
- \frac{1}{2} {\partial_{1} {\partial_{1} {\partial_{2} (\mathbf{u_{0}^{2}}) }}} + {\partial_{1} {\partial_{1} {\partial_{2} (u_{1}^{2}) }}} + {\partial_{1} {\partial_{1} {\partial_{2} (u_{2}^{2}) }}} + {\partial_{2} {\partial_{2} {\partial_{2} (u_{2}^{2}) }}} + {\partial_{1} {\partial_{1} {\partial_{1} (u_{1} u_{2}) }}} + 3 {\partial_{1} {\partial_{2} {\partial_{2} (u_{1} u_{2}) }}}\end{matrix}\right]
\label{eq:errorTerms3D_D3Q19S}
\end{equation}
\ifjournal \else \end{footnotesize} \fi
for the standard D3Q19-S model.
To illustrate the difference between these error terms, consider a duct flow in $x$-direction, where in the exact solution there are no transverse velocity components ($u_1=u_2=0$) i.e. only $u_0$ is different from zero. Only in the D3Q19-S model an error term containing $u_0$ enters the $y$ and $z$ second order error components, whereas in the D3Q27 and the D3Q19-I model there is no influence of $u_0$ on the $y$ and $z$
errors. The highlighted terms $\partial_1 \partial_2 \partial_2(u_0^2)$ in \eqref{eq:errorTerms3D_D3Q19S}, lead to spurious currents normal to the flow direction for the standard D3Q19-S model as is shown in \cite{Silva2014} and is confirmed by numerical experiments in the next section. Our theoretical results also correctly predict that in the D3Q27 and D3Q19-I models no such spurious currents occur.

\section{Numerical Experiments}
\label{sec:numerical_experiments}

In this section numerical experiments of two scenarios are presented: 
We first simulate a duct flow and evaluate spurious currents normal to the flow direction as in \cite{Silva2014} to confirm the theoretical results of the last section. As second test case, a nozzle geometry is investigated at Reynolds numbers where the standard D3Q19-S model is known to produce large deviations from the correct solution.
All simulations are conducted using the open source waLBerla\footnote{\url{http://www.walberla.net}} framework \cite{godenschwager2013framework}.
The framework was extended with an implementation of the improved D3Q19-I model, which has exactly the same number of memory accesses and floating point operations as the D3Q19-S model, and thus achieves the same performance.

\subsection{Poiseuille duct flow}

Many numerical tests of the Poiseuille duct flow have been conducted in the LB literature. Most of them focus on the velocity in flow direction and do not analyze velocity components normal to the flow direction. While these velocity components should be zero due to symmetry reasons, certain lattice models show spurious currents in these directions \cite{Silva2014}.
Here we repeat the numerical tests conducted by \cite{Silva2014} for a duct scenario with quadratic cross section at different grid resolutions. This is the simplest setup known to produce qualitatively different results for the D3Q19-S and D3Q27 models.

The scenario is set up with a square domain size of $(1,D,D)$, using periodic boundary conditions in $x$ direction. At the duct walls we employ half-way bounce back boundary conditions. All studies have also been run with full-way bounce back boundaries, however all results reported in the following are not altered by the chosen wall boundary condition. 
The channel is driven by a body force applied in $x$-direction using a forcing scheme with zero second order velocity moment \cite{Buick2000}. 
The duct diameter $D$ is varied between $15$ and $135$ cells while the relaxation rate is kept constant.
This diffusive scaling procedure coincides with the scaling employed in the theoretical analysis above ($\Delta x^2 = \Delta t$). 
The body force is chosen such that $Re=10$ is obtained. The flow is run until fully developed.

\begin{figure}[h]
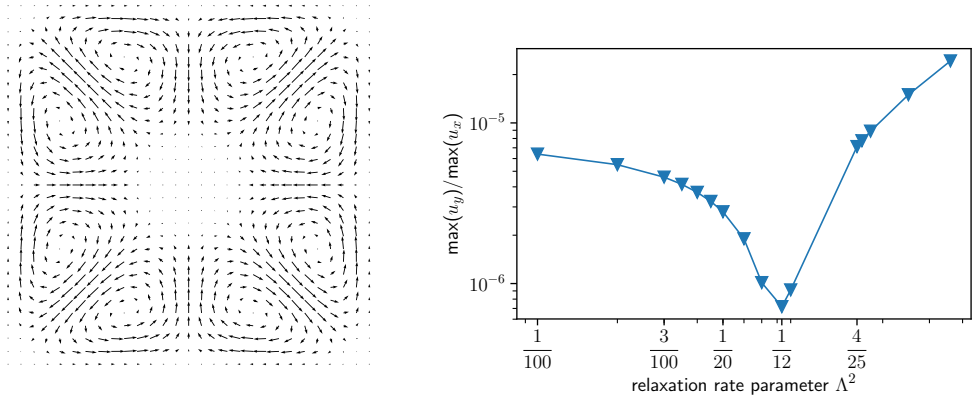

   \begin{center}
   \scalebox{0.45}{ \input{spurious_currents.pgf} }
   \scalebox{0.5}{ \input{lambda_dependency.pgf} }
   \end{center}

   \caption{Spurious currents in D3Q19-S duct flow. 
	     	Left: velocity field in a slice normal to flow direction.
		    Right: Dependency of spurious currents on relaxation parameter $\Lambda^2$ in D3Q19-S model. 
		    For D3Q19-I and D3Q27 the errors are in the range of machine precision for all values of $\Lambda^2$.}
   \label{fig:spurious_currents_D3Q19}
\end{figure}

As reported by \cite{Silva2014} and confirmed here, the D3Q19-S lattice model shows spurious currents in the $y\text{-}z$ plane (Fig.~\ref{fig:spurious_currents_D3Q19} left). We measure the size of the error using the maximum velocity, normal to the flow direction normalized with the maximum flow velocity: $\max(u_y)/\max(u_x)$. 
This quantity is dimensionless and therefore well suited for comparing results obtained at different grid resolutions.

\begin{figure}[h]
	\begin{center}
		\scalebox{0.8}{ \input{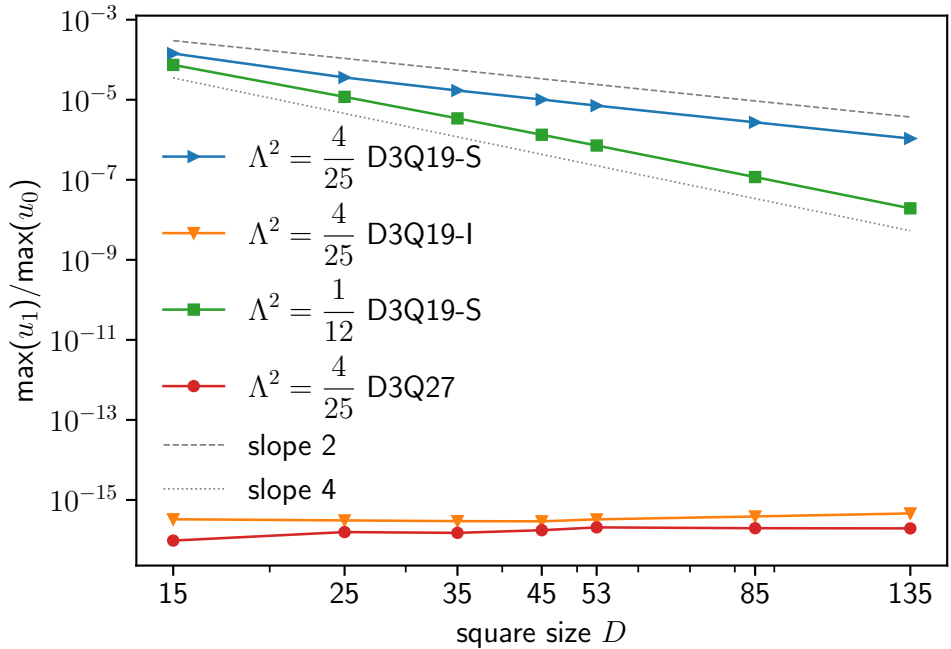} }
	\end{center}
	\caption{Spurious currents for different lattice models in duct flow scenario.}
	\label{fig:convergence_plot}
\end{figure}

The theoretical analysis shows that the spurious currents are caused by second order error terms of the momentum transport equation.
These terms are multiplied by a factor of $(\Lambda^2 - 1/12)$ and thus depend on the relaxation rate.
This dependency can be verified by numerical experiments as shown in Fig. \ref{fig:spurious_currents_D3Q19}. As predicted analytically, the minimum error is obtained at $\Lambda^2 = 1/12$. Fig. \ref{fig:convergence_plot} shows second order convergence of the error at non-optimal $\Lambda^2=4/25$. This confirms that the spurious currents are indeed caused by the second order error terms. If an optimal $\Lambda^2=1/12$ is chosen, the convergence is increased to fourth order. 
While being in perfect agreement with our theoretical analysis, the obtained convergence orders contradict the results from \cite{Silva2014}
who report convergence with third and fifth order respectively. This is most probably due to the fact that their error measure was not in non-dimensionalized form.

While the D3Q19-S model shows the discussed artifacts that converge to zero with increasing grid resolution, 
the D3Q27 model does not produce these spurious currents at all. It correctly predicts zero velocities normal to the flow direction up to machine accuracy, independent of the grid resolution and relaxation parameter. These results agree with the findings of \cite{Silva2014} and demonstrate the shortcomings of the D3Q19-S model. 
The improved D3Q19-I model, however, does not show any spurious currents and correctly predicts zero normal velocities, similar to the D3Q27 stencil. Thus, the observed numerical artifacts of D3Q19-S can not be caused by the reduced number of discrete velocities, but are a consequence of the discrete equilibrium choice.

\subsection{Nozzle Geometry}

\begin{figure}[h]
\begin{center}
\includegraphics[width=0.5\textwidth]{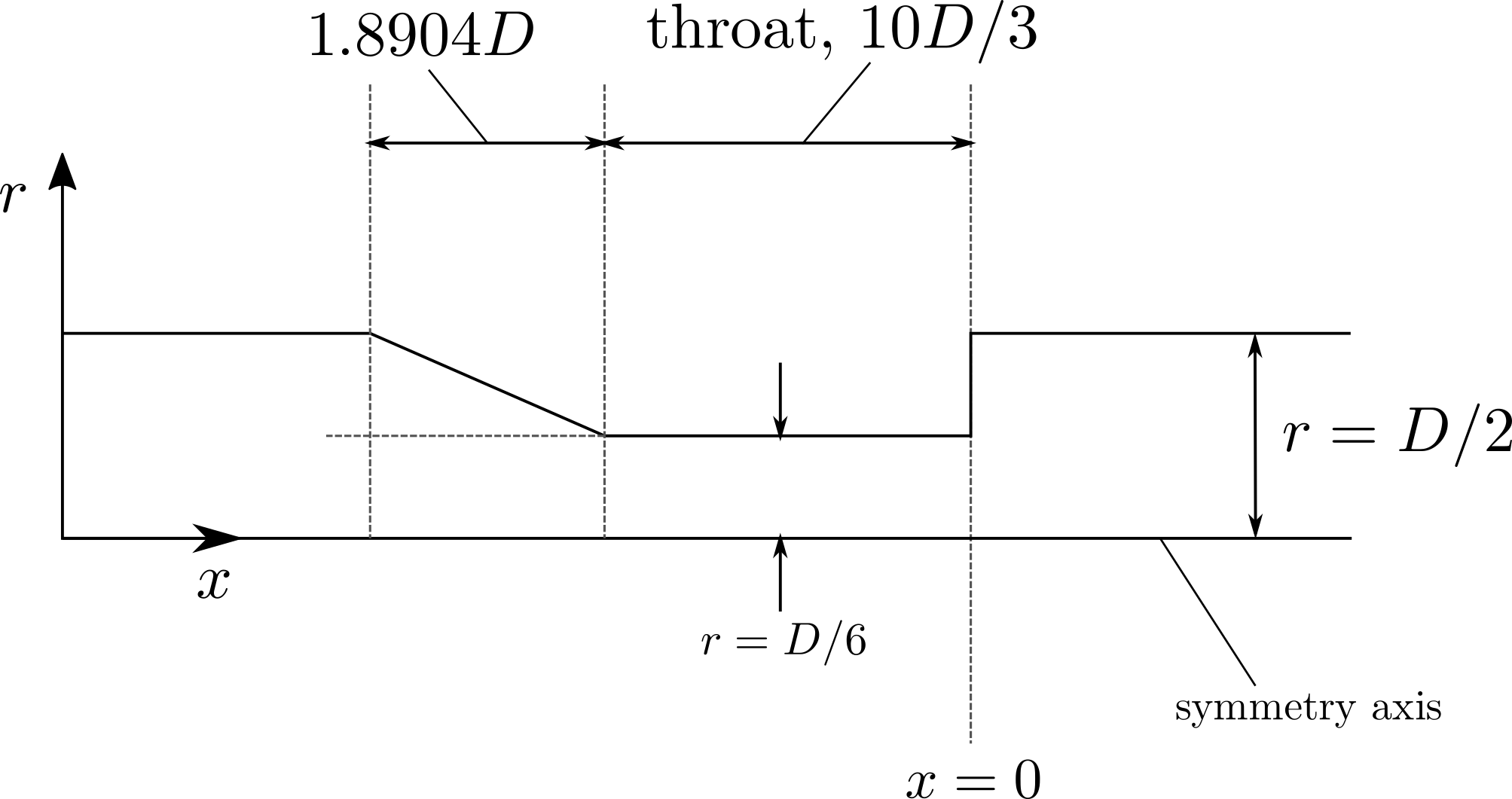}
\hspace{0.08\textwidth}
\includegraphics[width=0.4\textwidth]{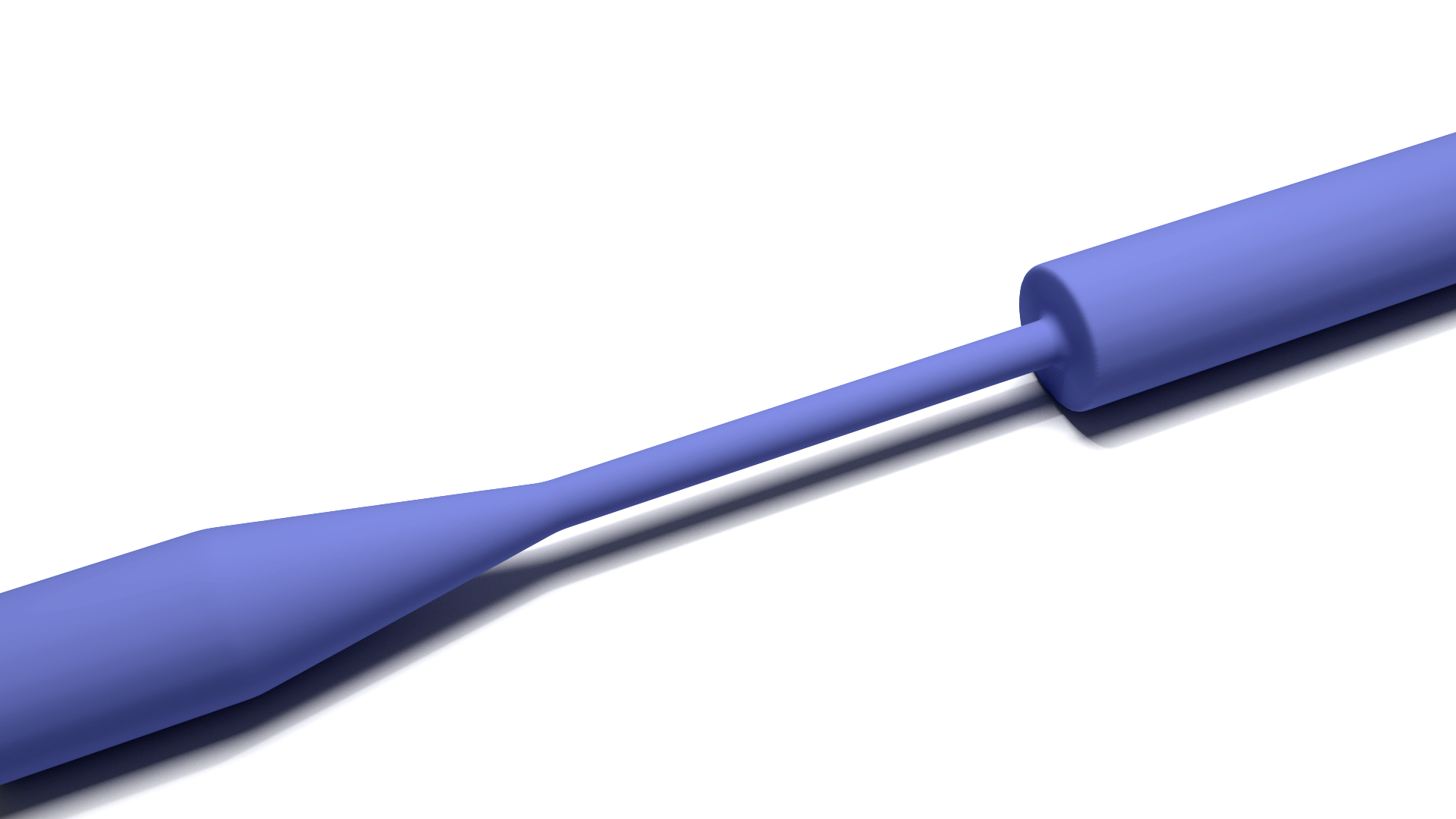}
\end{center}

\caption{Nozzle geometry for second simulation taken from \cite{White2011}: 
		 specifications - not to scale (left) and rendering (right)}
\label{fig:nozzle_geometry}
\end{figure}

As second test case we choose a nozzle geometry, which was employed by \cite{White2011} to investigate the 
deficiencies of the D3Q19 stencil compared to D3Q27. It is a benchmark flow model used by the US Food and Drug Administration (FDA) Critical Path Initiative Project \cite{hariharan2011multilaboratory}. 
As shown in Fig. \ref{fig:nozzle_geometry} this benchmark scenario consists of a pipe flow, where the pipe is linearly constricted to one third of its radius, leading to a ninefold increase in flow velocity at the throat. The throat then opens abruptly at $x=0$. The flow profile is examined in the plane $x=4D$, where a radially symmetric profile is expected for a laminar flow. For all lattice models two simulations are conducted at $Re=250$ and $Re=500$. As in \cite{White2011}, the Reynolds number is computed using the throat velocity and the reduced throat diameter $D/6$. 
For all simulations a pipe diameter $D=80$ is chosen. The simulation geometry is set up in waLBerla in parametric form using the Python interface of the framework~\cite{Bauer2016python}. At the inflow a parabolic velocity profile is set with a maximum velocity chosen such that the maximum lattice velocity at the throat is smaller than $0.07$. The outlet is chosen sufficiently long such that the flow is fully developed when it leaves the domain. As outflow boundary a constant pressure condition is employed.
It has been shown by \cite{White2011} that the D3Q19-S lattice model produces qualitatively different results than the D3Q27 model, especially for high Reynolds numbers. These deficiencies were found to be independent of grid density and Mach number, and thus suspected to be caused by the lack of isotropy of the D3Q19 stencil. We reproduce and confirm their results for the standard D3Q19-S and the D3Q27 model and, additionally, conduct the simulations also with our newly proposed D3Q19-I lattice model. The simulations are run until the flow is fully developed. As in \cite{White2011}, we simulate for $Re=250$ until  non-dimensional time $12$, and for $Re=500$ until $16.4$ in non-dimensional time defined as $t_d = D / u_{avg,in}$ with $D$ being the pipe diameter in cells and $u_{avg,in}$ the average inflow velocity in lattice units.

\begin{figure}[h]
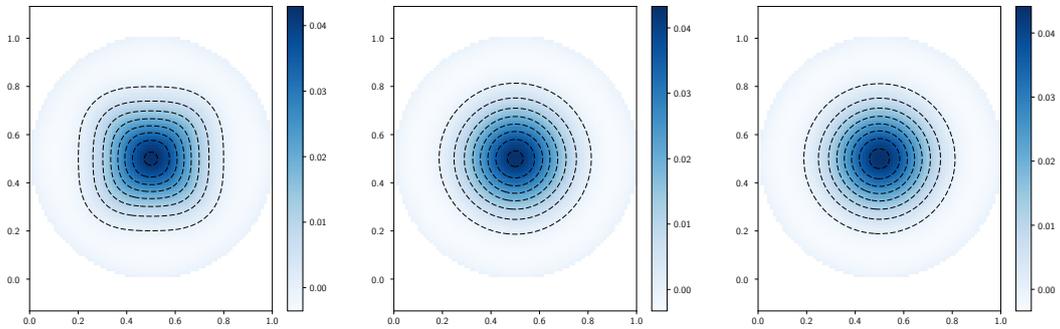

	\begin{center}
		\scalebox{0.36}{ \input{nozzle_D3Q19_old_Re250.pgf} }
		\scalebox{0.36}{ \input{nozzle_D3Q19_new_Re250.pgf} }
		\scalebox{0.36}{ \input{nozzle_D3Q27_Re250.pgf} }
	\end{center}
	\caption{Cross section at x=4D for D3Q19-S (left), D3Q19-I (middle) and D3Q27 (right) at $Re=250$ using simple BGK collision operator}
	\label{fig:nozzle_results_Re250}
\end{figure}

\begin{figure}[h]
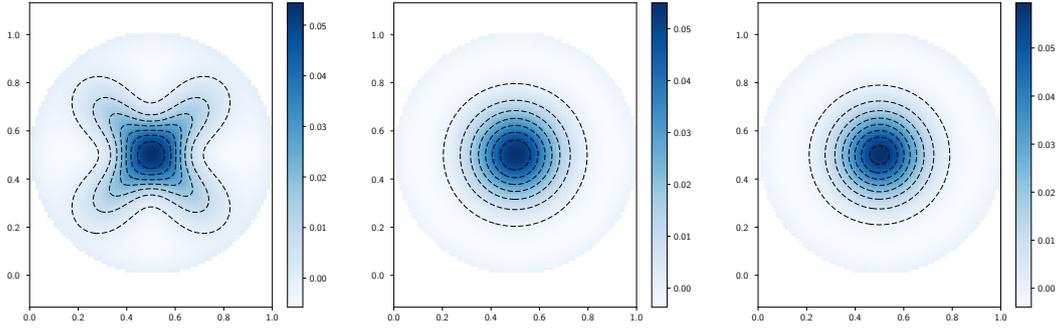

	\begin{center}
		\scalebox{0.36}{ \input{nozzle_D3Q19_old_Re500.pgf} }
		\scalebox{0.36}{ \input{nozzle_D3Q19_new_Re500.pgf} }
		\scalebox{0.36}{ \input{nozzle_D3Q27_Re500.pgf} }
	\end{center}
	\caption{Cross section at x=4D for D3Q19-S (left), D3Q19-I (middle) and D3Q27 (right) at $Re=500$
	         using entropically stabilized \texttt{KBC-N4} method}
	\label{fig:nozzle_results_Re500}
\end{figure}

Fig. \ref{fig:nozzle_results_Re250} shows contour plots of the velocity in flow direction $u_x$ at a slice
$x=4D$. Due to the symmetry of the geometry we expect the laminar flow profile to be radially symmetric. 
While the D3Q27 solution (right) shows perfect symmetry, the standard D3Q19 lattice model yields a qualitatively  different result. Despite requiring significantly less computational resources compared to the D3Q27 model, the improved D3Q19 model also correctly recovers the same radially symmetric solution.
%
The simulation at $Re=500$ with a simple BGK collision operator and the same grid resolution gets unstable due to a large relaxation rate of $1.978$. White et al. \cite{White2011} used a regularized BGK operator to overcome this problem. We instead use a more recently proposed, entropically stabilized MRT model similar to \texttt{KBC-N4} \cite{Bosch2015} where the relaxation rate for higher order moments is locally adapted according to a maximum entropy condition. Note that the relaxation of second order moments that control the fluid viscosity is not changed in this collision operator, so the numerical viscosity is still constant in the complete domain. Results of the $Re=500$ scenarios in Fig. \ref{fig:nozzle_results_Re500} show even larger artifacts for the D3Q19-S model which are fully compensated in the improved model. 

\section{Conclusion}

The D3Q19-S and D3Q27 lattice models both recover isothermal hydrodynamics in the limit of high
grid resolution and low Mach numbers. The D3Q19-S model, however shows anisotropic truncation errors
which are especially prominent in scenarios with high Reynolds numbers \cite{White2011}. 
We demonstrate different approaches to construct LB equilibria. The standard second order LB equilibrium 
is typically derived via a Hermite expansion approach that yields an expression for the equilibrium that is valid for all stencils and has to be adapted to concrete stencils by choosing different sets of weights.
We present an alternative equilibrium construction technique, based on the MRT idea of expressing the collision operator in moment space. To discretize the continuous Maxwellian distribution to a discrete velocity set, a linear independent set of equilibrium moments is chosen. The equilibrium values for these moments are then determined from the continuous Maxwellian. This leads to the same discrete equilibrium for D2Q9 and D3Q27 stencils, for D3Q19 a different, more accurate lattice model is obtained.
A steady state Chapman Enskog analysis shows that already the second order error terms of the two D3Q19 models differ. For D3Q19-S, the theoretical analysis predicts spurious currents normal to the flow direction in a duct, that decay quadratically with the grid resolution. The error terms causing these spurious currents are not present in our newly proposed D3Q19-I model.
Numerical experiments of the duct flow coincide with the theoretically predicted results. For this test case, the D3Q19-I model achieves similar accuracy as the D3Q27 version. The D3Q19 BGK simulation needs to load a factor of $19/27$ less data from memory and thus runs roughly $30\%$ faster, assuming well optimized, memory bound compute kernels and scenario sizes that do not fit into the outer level cache.
To summarize, we have presented an improved D3Q19 model that does not show the well-known accuracy deficiencies of the standard D3Q19-S model (compared to D3Q27) while keeping the superior runtime performance of the D3Q19-S model.

We have investigated two scenarios where the D3Q19-S model is known to produce qualitatively different results than D3Q27 and demonstrated that D3Q19-I does not show these anisotropic artifacts. What remains to be investigated is if the D3Q19-I model can achieve faster times to solution, i.e. if the runtime advantage is sufficient to outweigh the remaining accuracy gap.

\bibliographystyle{abbrvdin}
\bibliography{library}

\begin{thebibliography}{10}


\providecommand{\url}[1]{\texttt{#1}}
\expandafter\ifx\csname urlstyle\endcsname\relax
  \providecommand{\doi}[1]{doi: #1}\else
  \providecommand{\doi}{doi: \begingroup \urlstyle{rm}\Url}\fi

\bibitem[1]{Bauer2016python}
\textsc{Bauer}, M.  ; \textsc{Schornbaum}, F.  ; \textsc{Godenschwager}, C.  ;
  \textsc{Markl}, M.  ; \textsc{Anderl}, D.  ; \textsc{K{\"{o}}stler}, H.   ;
  \textsc{R{\"{u}}de}, U. :
\newblock {A Python extension for the massively parallel multiphysics
  simulation framework waLBerla}.
\newblock {In: }\emph{International Journal of Parallel, Emergent and
  Distributed Systems} 31 (2016), Nr. 6, 529--542.
\newblock  \url{http://dx.doi.org/10.1080/17445760.2015.1118478}. --
\newblock DOI 10.1080/17445760.2015.1118478

\bibitem[2]{Bosch2015}
\textsc{B{\"{o}}sch}, F.  ; \textsc{Chikatamarla}, S.~S.  ; \textsc{Karlin},
  I.~V.:
\newblock {Entropic multirelaxation lattice Boltzmann models for turbulent
  flows}.
\newblock {In: }\emph{Physical Review E - Statistical, Nonlinear, and Soft
  Matter Physics}  (2015).
\newblock  \url{http://dx.doi.org/10.1103/PhysRevE.92.043309}. --
\newblock DOI 10.1103/PhysRevE.92.043309. --
\newblock ISSN 15502376

\bibitem[3]{Buick2000}
\textsc{Buick}, J.~M. ; \textsc{Greated}, C.~A.:
\newblock {Gravity in a lattice Boltzmann model}.
\newblock {In: }\emph{Physical Review E} 61 (2000), Nr. 5, 5307--5320.
\newblock  \url{http://dx.doi.org/10.1103/PhysRevE.61.5307}. --
\newblock DOI 10.1103/PhysRevE.61.5307. --
\newblock ISBN 1063--651X

\bibitem[4]{Chapman1970}
\textsc{Chapman}, S.  ; \textsc{Cowling}:
\newblock \emph{{The mathematical theory of non-uniform gases: an account of
  the kinetic theory of viscosity, thermal conduction and diffusion in gases}}.
\newblock Cambridge University Press, 1970. --
\newblock  110--131 S.

\bibitem[5]{DHumieres2002}
\textsc{D'Humi{\`{e}}res}, D.  ; \textsc{Ginzburg}, I.  ; \textsc{Krafczyk}, M.
   ; \textsc{Lallemand}, P.   ; \textsc{Luo}, L.-S. :
\newblock {Multiple-relaxation-time lattice Boltzmann models in three
  dimensions.}
\newblock {In: }\emph{Philosophical transactions. Series A, Mathematical,
  physical, and engineering sciences} 360 (2002), Nr. 1792, S. 437--451.
\newblock  \url{http://dx.doi.org/10.1098/rsta.2001.0955}. --
\newblock DOI 10.1098/rsta.2001.0955. --
\newblock ISBN 1364503X

\bibitem[6]{frisch1987lattice}
\textsc{Frisch}, U.  ; \textsc{Hasslacher}, B.   ; \textsc{Lallemand}, P. :
\newblock {Lattice Gas Hydrodynamics in Two and Three Dimensions}.
\newblock {In: }\emph{Complex Systems} 1 (1987), S. 649--707

\bibitem[7]{ginzburg2008consistent}
\textsc{Ginzburg}, I. :
\newblock {Consistent lattice Boltzmann schemes for the Brinkman model of
  porous flow and infinite Chapman-Enskog expansion}.
\newblock {In: }\emph{Physical Review E} 77 (2008), Nr. 6, S. 66704

\bibitem[8]{godenschwager2013framework}
\textsc{Godenschwager}, C.  ; \textsc{Schornbaum}, F.  ; \textsc{Bauer}, M.  ;
  \textsc{K{\"{o}}stler}, H.   ; \textsc{R{\"{u}}de}, U. :
\newblock {A framework for hybrid parallel flow simulations with a trillion
  cells in complex geometries}.
\newblock {In: }\emph{Proceedings of the International Conference on High
  Performance Computing, Networking, Storage and Analysis} ACM, 2013, S.~35

\bibitem[9]{Guo2002}
\textsc{Guo}, Z.  ; \textsc{Zheng}, C.   ; \textsc{Shi}, B. :
\newblock {Discrete lattice effects on the forcing term in the lattice
  Boltzmann method}.
\newblock {In: }\emph{Physical Review E - Statistical, Nonlinear, and Soft
  Matter Physics} 65 (2002), Nr. 4, S. 1--6.
\newblock  \url{http://dx.doi.org/10.1103/PhysRevE.65.046308}. --
\newblock DOI 10.1103/PhysRevE.65.046308. --
\newblock ISBN 1063--651X

\bibitem[10]{hariharan2011multilaboratory}
\textsc{Hariharan}, P.  ; \textsc{Giarra}, M.  ; \textsc{Reddy}, V.  ;
  \textsc{Day}, S.~W. ; \textsc{Manning}, K.~B. ; \textsc{Deutsch}, S.  ;
  \textsc{Stewart}, S. F.~C. ; \textsc{Myers}, M.~R. ; \textsc{Berman}, M.~R. ;
  \textsc{Burgreen}, G.~W.  ; \textsc{Others}:
\newblock {Multilaboratory particle image velocimetry analysis of the FDA
  benchmark nozzle model to support validation of computational fluid dynamics
  simulations}.
\newblock {In: }\emph{Journal of biomechanical engineering} 133 (2011), Nr. 4,
  S. 41002

\bibitem[11]{harrison2007use}
\textsc{Harrison}, S.~E.:
\newblock \emph{{The use of the lattice Boltzmann method in thrombosis
  modelling.}}, University of Sheffield, Diss., 2007

\bibitem[12]{He1997}
\textsc{He}, X.  ; \textsc{Luo}, L.-S. :
\newblock {Theory of the lattice Boltzmann method: From the Boltzmann equation
  to the lattice Boltzmann equation}.
\newblock {In: }\emph{Physical Review E} 56 (1997), Nr. 6, S. 6811--6817.
\newblock  \url{http://dx.doi.org/10.1103/PhysRevE.56.6811}. --
\newblock DOI 10.1103/PhysRevE.56.6811. --
\newblock ISBN 1063--651X

\bibitem[13]{he1997theory}
\textsc{He}, X.  ; \textsc{Luo}, L.-S. :
\newblock {Theory of the lattice Boltzmann method: From the Boltzmann equation
  to the lattice Boltzmann equation}.
\newblock {In: }\emph{Physical Review E} 56 (1997), Nr. 6, S. 6811

\bibitem[14]{kang2013effect}
\textsc{Kang}, S.~K. ; \textsc{Hassan}, Y.~A.:
\newblock {The effect of lattice models within the lattice Boltzmann method in
  the simulation of wall-bounded turbulent flows}.
\newblock {In: }\emph{Journal of Computational Physics} 232 (2013), Nr. 1, S.
  100--117

\bibitem[15]{Khirevich2015}
\textsc{Khirevich}, S.  ; \textsc{Ginzburg}, I.   ; \textsc{Tallarek}, U. :
\newblock {Coarse-and fine-grid numerical behavior of MRT/TRT lattice-Boltzmann
  schemes in regular and random sphere packings}.
\newblock {In: }\emph{Journal of Computational Physics}  (2015).
\newblock  \url{http://dx.doi.org/10.1016/j.jcp.2014.10.038}. --
\newblock DOI 10.1016/j.jcp.2014.10.038. --
\newblock ISSN 10902716

\bibitem[16]{Kruger2016}
\textsc{Kr{\"{u}}ger}, T.  ; \textsc{Kusumaatmaja}, H.  ; \textsc{Kuzmin}, A.
  ; \textsc{Shardt}, O.  ; \textsc{Silva}, G.   ; \textsc{Viggen}, E.~M.:
\newblock \emph{{The Lattice Bolzmann Method - Principles and Practics}}.
\newblock Springer, 2017. --
\newblock ISBN 978--3--319--44649--3

\bibitem[17]{Lallemand2000}
\textsc{Lallemand}, P.  ; \textsc{Luo}, L.-s. :
\newblock {Theory of the lattice Boltzmann method: Dispersion, dissipation,
  isotropy,Galilean invariance, and stability}.
\newblock {In: }\emph{Physical Review E} 61 (2000), Nr. 6, 6546--6562.
\newblock  \url{http://dx.doi.org/10.1103/PhysRevE.61.6546}. --
\newblock DOI 10.1103/PhysRevE.61.6546. --
\newblock ISBN 1063--651X

\bibitem[18]{mayer2006direct}
\textsc{Mayer}, G.  ; \textsc{H{\'{a}}zi}, G. :
\newblock {Direct numerical and large eddy simulation of longitudinal flow
  along triangular array of rods using the lattice Boltzmann method}.
\newblock {In: }\emph{Mathematics and Computers in Simulation} 72 (2006), Nr.
  2, S. 173--178

\bibitem[19]{Shan1998}
\textsc{Shan}, X.  ; \textsc{He}, X. :
\newblock {Discretization of the Velocity Space in the Solution of the
  Boltzmann Equation}.
\newblock {In: }\emph{Physical Review Letters} 80 (1998), Nr. 1, 65--68.
\newblock  \url{http://dx.doi.org/10.1103/PhysRevLett.80.65}. --
\newblock DOI 10.1103/PhysRevLett.80.65. --
\newblock ISSN 0031--9007

\bibitem[20]{Shan2006}
\textsc{Shan}, X.  ; \textsc{Yuan}, X.-F.   ; \textsc{Chen}, H. :
\newblock {Kinetic theory representation of hydrodynamics: a way beyond the
  Navier–Stokes equation}.
\newblock {In: }\emph{Journal of Fluid Mechanics} 550 (2006), Nr. 1, S. 413.
\newblock  \url{http://dx.doi.org/10.1017/S0022112005008153}. --
\newblock DOI 10.1017/S0022112005008153. --
\newblock ISBN 0022--1120

\bibitem[21]{Silva2011}
\textsc{Silva}, G.  ; \textsc{Semiao}, V. :
\newblock {A study on the inclusion of body forces in the lattice Boltzmann BGK
  equation to recover steady-state hydrodynamics}.
\newblock {In: }\emph{Physica A: Statistical Mechanics and its Applications}
  390 (2011), Nr. 6, 1085--1095.
\newblock  \url{http://dx.doi.org/10.1016/j.physa.2010.11.037}. --
\newblock DOI 10.1016/j.physa.2010.11.037. --
\newblock ISBN 0378--4371

\bibitem[22]{Silva2014}
\textsc{Silva}, G.  ; \textsc{Semiao}, V. :
\newblock {Truncation errors and the rotational invariance of three-dimensional
  lattice models in the lattice Boltzmann method}.
\newblock {In: }\emph{Journal of Computational Physics}  (2014).
\newblock  \url{http://dx.doi.org/10.1016/j.jcp.2014.03.027}. --
\newblock DOI 10.1016/j.jcp.2014.03.027. --
\newblock ISBN 0021--9991

\bibitem[23]{Stiebler2011}
\textsc{Stiebler}, M.  ; \textsc{Freudiger}, S.  ; \textsc{Krafczyk}, M.   ;
  \textsc{Geier}, M. :
\newblock {Parallel Lattice-Boltzmann simulation of transitional flow on
  non-uniform grids}.
\newblock {In: }\emph{Notes on Numerical Fluid Mechanics and Multidisciplinary
  Design}, 2011. --
\newblock ISBN 9783642177699

\bibitem[24]{Tolke2006}
\textsc{T{\"{o}}lke}, J.  ; \textsc{Freudiger}, S.   ; \textsc{Krafczyk}, M. :
\newblock {An adaptive scheme using hierarchical grids for lattice Boltzmann
  multi-phase flow simulations}.
\newblock {In: }\emph{Computers and Fluids}  (2006).
\newblock  \url{http://dx.doi.org/10.1016/j.compfluid.2005.08.010}. --
\newblock DOI 10.1016/j.compfluid.2005.08.010. --
\newblock ISBN 0045--7930

\bibitem[25]{Wellein2006}
\textsc{Wellein}, G.  ; \textsc{Zeiser}, T.  ; \textsc{Hager}, G.   ;
  \textsc{Donath}, S. :
\newblock {On the single processor performance of simple lattice Boltzmann
  kernels}.
\newblock {In: }\emph{Computers and Fluids} 35 (2006), Nr. 8-9, S. 910--919.
\newblock  \url{http://dx.doi.org/10.1016/j.compfluid.2005.02.008}. --
\newblock DOI 10.1016/j.compfluid.2005.02.008. --
\newblock ISBN 0045--7930

\bibitem[26]{White2011}
\textsc{White}, A.~T. ; \textsc{Chong}, C.~K.:
\newblock {Rotational invariance in the three-dimensional lattice Boltzmann
  method is dependent on the choice of lattice}.
\newblock {In: }\emph{Journal of Computational Physics}  (2011).
\newblock  \url{http://dx.doi.org/10.1016/j.jcp.2011.04.031}. --
\newblock DOI 10.1016/j.jcp.2011.04.031. --
\newblock ISBN 0021--9991

\bibitem[27]{Wittmann2013a}
\textsc{Wittmann}, M.  ; \textsc{Zeiser}, T.  ; \textsc{Hager}, G.   ;
  \textsc{Wellein}, G. :
\newblock {Comparison of different propagation steps for lattice Boltzmann
  methods}.
\newblock {In: }\emph{Computers and Mathematics with Applications} 65 (2013),
  Nr. 6, 924--935.
\newblock  \url{http://dx.doi.org/10.1016/j.camwa.2012.05.002}. --
\newblock DOI 10.1016/j.camwa.2012.05.002. --
\newblock ISBN 0898--1221

\bibitem[28]{zou1995improved}
\textsc{Zou}, Q.  ; \textsc{Hou}, S.  ; \textsc{Chen}, S.   ; \textsc{Doolen},
  G.~D.:
\newblock {A improved incompressible lattice Boltzmann model for
  time-independent flows}.
\newblock {In: }\emph{Journal of Statistical Physics} 81 (1995), Nr. 1, S.
  35--48

\end{thebibliography}

\end{document}